\documentclass[aps,prd,twocolumn,showpacs,superscriptaddress, nofootinbib]{revtex4-2}
\usepackage{amsmath,amssymb,graphicx,hyperref}
\usepackage{caption}
\usepackage{float}

\usepackage{xcolor}
\usepackage{amsmath}
\definecolor{linkblack}{RGB}{20,20,20}
\definecolor{citeblue}{RGB}{0,70,140}
\definecolor{urlblue}{RGB}{0,90,120}
\usepackage{float}
\hypersetup{
  colorlinks=true,
  linkcolor=linkblack,
  citecolor=citeblue,
  urlcolor=urlblue
}

\DeclareUnicodeCharacter{2212}{-}

\DeclareMathOperator{\sech}{sech}

\newcommand{\eq}[1]{Eq.~\ref{#1}}

\newcommand{\figs}[2]{Figs.~(\ref{#1}) and~(\ref{#2})}

\begin{document}

\title{Open case for a closed universe}

\author{Nathan L. Burwig}
\affiliation{Department of Physics \& Beyond Center for Fundamental Concepts in Science, Arizona State University, Tempe, AZ 85287-1504, USA}
\author{Damien A. Easson}
\affiliation{Department of Physics \& Beyond Center for Fundamental Concepts in Science, Arizona State University, Tempe, AZ 85287-1504, USA}


\begin{abstract}
We establish a new no-go theorem for cosmology: spatially flat ($k=0$) and open ($k=-1$) Friedmann–Robertson–Walker (FRW) non-static spacetimes cannot be simultaneously nonsingular, geodesically complete, and consistent with the averaged null energy condition (ANEC). Equivalently, any dynamic flat or open universe that is complete must violate the ANEC. By contrast, closed universes ($k=+1$) uniquely admit nonsingular, geodesically complete, ANEC-consistent solutions, with global de Sitter space as the canonical realization that saturates the ANEC. Furthermore, we analytically demonstrate that positive spatial curvature naturally mimics the phenomenology of phantom dark energy ($w<-1$), biasing flat-model reconstructions of $w(z)$ at the $\sim 1\%$ level. These results augment the classical singularity theorems, establish a new classification of eternal cosmologies, and motivate renewed scrutiny of spatial curvature in both theory and observation.
\end{abstract}

\maketitle

\section{Introduction}
The singularity theorems of Hawking and Penrose demonstrate that, under broad and physically reasonable assumptions, including the validity of classical energy conditions and the presence of trapped surfaces, generic spacetimes in general relativity are geodesically incomplete~\cite{Penrose:1964wq,Hawking:1966sx,Hawking:1966jv,Hawking:1967ju,Hawking:1970zqf,Hawking:1973uf}. These results are often interpreted as suggesting that cosmological spacetimes contain a past singularity, such as a Big Bang singularity, although the theorems technically establish only the existence of incomplete geodesics, and not necessarily a singularity in curvature or energy density. 

A related but distinct result is the Borde--Guth--Vilenkin (BGV) theorem, which addresses the past completeness of inflationary spacetimes  \cite{Borde:2001nh}. The BGV theorem claims that any universe which has, on average, been expanding along a past-directed geodesic must be geodesically incomplete to the past, regardless of the details of the energy content or the validity of classical energy conditions. Unlike the Hawking–Penrose theorems, the BGV result does not require the presence of trapped surfaces or any specific matter model; it relies only on the assumption of positive average Hubble expansion along timelike or null geodesics. Consequently, the theorem is often interpreted as suggesting that inflationary spacetimes, even if eternal into the future, cannot be extended indefinitely into the past within a classical spacetime description \cite{Kinney:2023urn,Geshnizjani:2023hyd}. The theorem, however, admits certain loopholes, and examples of past-eternal inflationary models are prevalent~\cite{Ellis:2002we, Ellis:2003qz, Lesnefsky:2022fen, Easson:2024uxe, Easson:2024fzn}. 

While the point-wise standard energy conditions play a key role in the singularity theorems of GR, the averaged null energy condition (ANEC) occupies a special place at the intersection of quantum field theory, gravitation, and cosmology \cite{Wald:1991xn, Fewster:2006uf, Graham:2007va, Wall:2009wi, Hartman:2016lgu}. It asserts that the integral of the null--null component of the stress--energy tensor along any complete null geodesic is non-negative,
\begin{equation}
\int_{-\infty}^{+\infty} T_{\mu\nu} k^\mu k^\nu \, d\lambda \ge 0.
\end{equation}
Unlike the classical pointwise energy conditions, which are routinely violated by quantum fluctuations, the ANEC captures the weaker but physically robust statement that \emph{negative energy densities cannot persist without compensation}. It embodies the idea that, although quantum fields may locally exhibit negative energy, the total energy flux measured along a light ray must remain positive in any consistent, unitary theory.

From a cosmological perspective, the ANEC serves as a minimal energy-positivity criterion that any physically acceptable universe should satisfy on average. It ensures that light rays are, in aggregate, gravitationally focused rather than defocused, enforcing causal structure and protecting the global coherence of spacetime. If the ANEC were grossly violated, the universe could admit acausal shortcuts, cyclic paradoxes, or uncontrolled instabilities. Conversely, adherence to the ANEC provides a natural demarcation between physically reasonable and exotic cosmologies. Our analysis therefore treats the ANEC not as an optional assumption but as a fundamental consistency condition that any realistic cosmological spacetime, including our own, must satisfy.

A central question is whether fully nonsingular cosmologies can exist without invoking exotic matter or abandoning robust principles such as the ANEC.

In this paper we establish a new theorem:
\emph{non-static, spatially flat FRW spacetimes cannot be simultaneously
nonsingular, geodesically complete, and consistent with the averaged null
energy condition (ANEC).}
Equivalently,
\emph{any non-static flat FRW universe that satisfies the ANEC is necessarily
geodesically incomplete.}
There is exactly one way to reconcile flatness, regularity, and completeness---through ANEC violation.
By contrast, only closed ($k=+1$) universes can satisfy geodesic completeness
and the ANEC simultaneously, with global de~Sitter space providing the
canonical realization that saturates the bound.
Further details are relegated to our companion paper \cite{burwig2:2025dae}.
We use natural units with $\hbar=c_{light}=1$ and express all quantities in reduced Planck units by setting 
the Planck mass $M_{pl}= 1/\sqrt{8 \pi G}$=1.

\section{Main Theorem}
Consider the FRW line element
\begin{equation}
ds^{2} = -dt^{2} + a(t)^{2}\, d\Sigma_{k}^{2}, \qquad k\in\{0,\pm1\},
\end{equation}
with $a\in C^{2}$ and $a(t)>0$. For a perfect fluid source we define the affine ANEC integral
via two-sided improper truncations
\begin{equation}
I_{\mathrm{ANEC}}
\;:=\;
\lim_{T_\pm\to \pm \infty}\,\int_{T_-}^{T_+}\frac{\rho(t)+p(t)}{a(t)}\,dt \,.
\label{eq:ANEC-def}
\end{equation}
We refrain from separating boundary and bulk limits unless each converges individually.\footnote{For radial null geodesics in an FRW spacetime, enforcing affine
parametrization of the null tangent vector yields
$dt/d\lambda = 1/a(t)$, up to an arbitrary positive constant that simply
rescales the affine parameter. Equivalently, one may take
$d\lambda = a(t)\,dt$. This choice is standard, unique up to rescaling,
and does not affect the sign or validity of the averaged null energy
condition.}

\medskip
\noindent \textbf{Theorem (Flat/Open FRW).}
Let $(M,g)$ be a non-static FRW spacetime with $k=0$ or $k=-1$, $a(t)\in C^2$, $a(t)>0$, and bounded curvature invariants. If $(M,g)$ is null geodesically complete, then
\begin{equation}
I_{\mathrm{ANEC}} < 0 .
\end{equation}
Hence, any ANEC-satisfying ($I_{\mathrm{ANEC}}\geq 0$) flat or open FRW spacetime is null-incomplete.

\emph{Proof sketch--}Work on a finite interval $[T_-,T_+]$ and then pass to the two-sided improper limit \eqref{eq:ANEC-def}.
For $k=0$ the finite-interval identity is
\begin{equation}
\int_{T_-}^{T_+}\frac{\rho+p}{a}\,dt
= -\,2\Big[\tfrac{H}{a}\Big]_{T_-}^{T_+}
\;-\;2\int_{T_-}^{T_+}\frac{H^2}{a}\,dt,
\label{eq:flat-identity}
\end{equation}
where $H:=\dot a/{a}$ is the Hubble parameter.\\

\noindent\emph{Claim (boundary control).}
Under bounded curvature and null completeness, either
$\displaystyle \int H^{2}/a\,dt=+\infty$ along the truncations (so $I_{\mathrm{ANEC}}=-\infty$),
or else $\displaystyle \int H^{2}/a\,dt<\infty$, in which case
$\displaystyle [H/a]_{T_-}^{T_+}\to0$ along two-sided truncations.~\footnote{In the finite-bulk branch, bounded curvature implies $H$ is uniformly continuous;
together with $\int H^{2}/a\,dt<\infty$ this forces $H/a\to0$, so the boundary term
vanishes.}

\noindent By \eqref{eq:flat-identity}, in the finite-bulk case the boundary term vanishes and
\[
I_{\mathrm{ANEC}}
=\,-2\!\int_{-\infty}^{\infty}\!\frac{H^2}{a}\,dt \;<\;0,
\]
since $H^2/a\ge0$ and is strictly positive on a set of nonzero measure for any non-static model ($H\not\equiv0$).
For $k=-1$ the corresponding finite-interval identity contains an additional \emph{negative} curvature contribution in the bulk integral
\begin{equation}
\int_{T_-}^{T_+}\!\frac{\rho+p}{a}\,dt
 = -2\!\!\int_{T_-}^{T_+}\!\!\frac{H^2+1/a^2}{a}\,dt
   -2\!\!\left[\frac{H}{a}\right]_{T_-}^{T_+},
\end{equation}
so the same conclusion holds \emph{a fortiori}: null completeness forces $I_{\mathrm{ANEC}}<0$ in the open case as well. \hfill$\square$

\medskip
\noindent \textbf{Closed Case ($k=+1$).}
We adopt \emph{symmetric} truncations $[-T,T]$, which respect the natural time-reflection symmetry of nonsingular closed cosmologies (including global de~Sitter) and ensure that boundary terms in the affine ANEC integral are treated on equal footing in the two-sided limit. On $[-T,T]$ the geometric identity is
\begin{equation}
I_{\mathrm{ANEC}}(T)
= -\,2\Big[\tfrac{H}{a}\Big]_{-T}^{T}
\;+\; 2\!\int_{-T}^{T}\!\Big(\tfrac{1}{a^{3}}-\tfrac{H^{2}}{a}\Big)\,dt,
\label{eq:k1-finite}
\end{equation}
and the affine ANEC is the improper two-sided limit
\begin{equation}
I_{\mathrm{ANEC}} := \lim_{T\to\infty} I_{\mathrm{ANEC}}(T).
\label{eq:k1-limit}
\end{equation}
Under bounded curvature, $H$ is bounded and $a$ is bounded away from $0$, hence $[H/a]_{-T}^{T}$ is \emph{uniformly bounded} along the truncations. Consequently:
\begin{itemize}
\item[(a)] If $\displaystyle\int_{-\infty}^{\infty}\! a^{-3}\,dt=+\infty$ and $\displaystyle\int_{-\infty}^{\infty}\! H^{2}/a\,dt<\infty$, then $I_{\mathrm{ANEC}}=+\infty$.
\item[(b)] If $\displaystyle\int_{-\infty}^{\infty}\! a^{-3}\,dt<\infty$ and $\displaystyle\int_{-\infty}^{\infty}\! H^{2}/a\,dt<\infty$, then $I_{\mathrm{ANEC}}$ is finite.
\item[(c)] If $\displaystyle\int_{-\infty}^{\infty}\! H^{2}/a\,dt=+\infty$ and $\displaystyle\int_{-\infty}^{\infty}\! a^{-3}\,dt<\infty$, then $I_{\mathrm{ANEC}}=-\infty$.
\item[(d)] If both $\displaystyle\int a^{-3}\,dt$ and $\displaystyle\int H^{2}/a\,dt$ diverge, the sign is governed by the competition between the divergent bulk integrals in \eqref{eq:k1-finite}; the bounded boundary term becomes subdominant and does not affect the asymptotic sign.
\end{itemize}

A sufficient (though not necessary) condition for the
boundary term in Eq.~\eqref{eq:k1-finite} to vanish is that the scale factor
satisfy $\int_{-\infty}^{\infty} a^{-1}(t)\,dt<\infty$
(e.g.\ exponential or faster growth on both ends).
For nonsingular FRW spacetimes (i.e.\ with bounded curvature invariants),
the Hubble parameter $H$ is necessarily bounded.
Hence $\int_{-\infty}^{\infty} (H^2/a)\,dt<\infty$ and the boundary
term vanishes, so
\begin{equation}
I_{\mathrm{ANEC}} \;=\; 2\!\int_{-\infty}^{\infty}\!
\Big(\frac{1}{a^{3}}-\frac{H^{2}}{a}\Big)\,dt \,.
\label{eq:IANEC_k1}
\end{equation}
In particular, whenever the boundary term vanishes and $\int (H^2/a)\,dt<\infty$,
\begin{equation}
I_{\mathrm{ANEC}}=+\infty
\;\;\Longleftrightarrow\;\;
\int_{-\infty}^{\infty}\frac{dt}{a^{3}(t)}=+\infty.
\label{eq:IANEC_div}
\end{equation}

The canonical example is global de Sitter with $a(t)=h^{-1}\cosh(ht)$, for which $\rho+p=0$ and $I_{\mathrm{ANEC}}=0$ (saturation).
\footnote{The flat/open implication is strict (\(I_{\rm ANEC}<0\) under completeness) with the unique exception of Minkowski, whereas in the closed case no universal sign holds; saturation \(I_{\rm ANEC}=0\) occurs in global de Sitter, as discussed below.}

\section{Illustrative examples}
\subsection{Flat cosh bounce} Consider $a(t)=h^{-1}\cosh(ht)$ in a $k=0$ FRW universe. This gives a nonsingular bouncing toy cosmology, in which the universe contracts to a minimum size and then expands. The model is inflating for all time: $\ddot a = h\cosh(ht)>0$.  The spacetime is geodesically complete, but the null energy condition is violated everywhere: $\rho+p=-2h^{2}\mathrm{sech}^{2}(ht)<0$, and the ANEC integral is 
\begin{equation}
I_{\text{ANEC}}
 = -2h^2\!\!\int_{-\infty}^{\infty}\!\!\frac{\sech^2(ht)}{a(t)}\,dt
 = -\pi h^2 \,.
\end{equation}
The supporting matter is phantom-like, violating all classical energy conditions.

\subsection{Closed cosh bounce}  
The same scale factor in a $k=+1$ FRW spacetime corresponds to global de Sitter. Here $\rho+p=0$ identically, so the NEC and ANEC are saturated: $I_{\text{ANEC}}=0$. The model is maximally symmetric, nonsingular, and supported by a positive cosmological constant. Thus pure de Sitter is a stark illustration of the theorem and serves as a simple example of a nonsingular curvature-supported bouncing cosmology which satisfies all but the strong energy condition; although, it is at best a toy cosmological model as it is supported by a constant energy density, and contains no matter.

\subsection{Past-eternal inflationary model with exit} 
Consider the scale factor
\begin{equation}\label{atanh}
a(t)=a_0\sqrt{\tanh\!\big(t/\alpha\big)+c},\qquad \alpha>0,\;c>1,
\end{equation}
which leads to a nonsingular FRW cosmology consistent with the completeness theorem of \cite{Lesnefsky:2022fen}. The constant $c$ controls the nonsingular early-time behavior. For $c > 1$, \eq{atanh} remains strictly positive for all $t$, ensuring geodesic completeness and avoiding singularities. As $t \to -\infty$, the scale factor asymptotes to a finite minimum value $a_{\min} = a_0 \sqrt{c - 1}$, corresponding to an emergent past of asymptotic constant size. This is not a finite-time bounce but a \emph{bounce at infinity} in the sense of \cite{Easson:2024fzn}: the universe is eternally expanding and nonsingular in both time directions. The parameter $c$ thus regulates the depth of the asymptotic phase and the degree of early-time curvature support.

The function in \eq{atanh} is smooth and strictly increasing for all $t$, with finite limits $a(\pm\infty)=a_0\sqrt{c\pm 1}$. Writing $y\equiv\tanh(t/\alpha)$ and $f\equiv\sech^2(t/\alpha)=1-y^2$, one finds
\begin{equation}
H=\frac{\dot a}{a}=\frac{f}{2\alpha(y+c)}, 
\end{equation}
\begin{equation}\label{hdot}
    \dot H=-\frac{f}{2\alpha^2(y+c)^2}\,\Big(2y(y+c)+f\Big).
\end{equation}
The standard first Hubble flow parameter is 
\begin{equation}
\epsilon_1\equiv-\frac{\dot H}{H^2}=2+\frac{4y(y+c)}{1-y^2},
\end{equation}
so accelerated, inflationary expansion ($\ddot a>0$) occurs when $3y^2+4cy+1<0$, i.e. for $y\in(-1,y_+)$ with $y_+=(-2c+\sqrt{4c^2-3})/3\in(-1,0)$. Thus the model has a past-eternal inflationary phase which smoothly self‑terminates at $t_{\rm exit}=\alpha\,\mathrm{arctanh}(y_+)$, after which $\ddot a<0$ and the strong energy condition (SEC) is restored.

\begin{figure}[H]
\centering
  \includegraphics[width=1\linewidth]{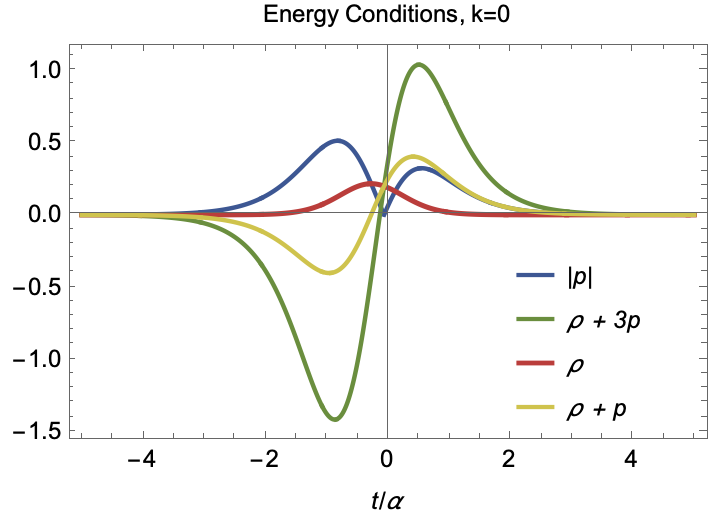}
  \caption{Energy conditions from \eq{atanh} with $k=0$. Plot of energy density $\rho$ (red), $\rho + p$  (yellow), $|p|$ (blue) and $\rho + 3p$ (green).}
  \label{ECtanhk0}
\end{figure}

Energy conditions distinguish the flat and closed cases (See \figs{ECtanhk0}{ECtanhk1}, respectively). For $k=0$ one has $\rho+p=-2\dot H$, which is negative when $\dot H>0$; from \eq{hdot}, $\dot H>0$ for an infinite past interval, so the NEC is violated for some time.

The ANEC: 
\begin{equation}
I_{\rm ANEC}^{(k=0)}
= \frac{4\!\left(\sqrt{c-1}+\sqrt{c+1}
+ 2c\big(\sqrt{c-1}-\sqrt{c+1}\big)\right)}
{3\,\alpha\,a_0\,\sqrt{c^{2}-1}}\,,
\end{equation}
is negative for $c>1$, and hence, violated.

In contrast, for $k=+1$, $I_{\rm ANEC}^{(k=1)} = + \infty$, so the ANEC is satisfied in the strongest possible sense.
Hence the closed model satisfies NEC, WEC and DEC at all times (and hence the ANEC), while violating only the SEC during the finite inflationary interval before $t_{\rm exit}$. Unlike the de Sitter bounce, the present solution is not eternally accelerating: it is a nonsingular, geodesically complete example of curvature‑powered \emph{inflation with graceful exit} realized with standard matter in GR, without invoking exotic fields or modified gravity.

\begin{figure}[H]
\centering
  \includegraphics[width=1\linewidth]{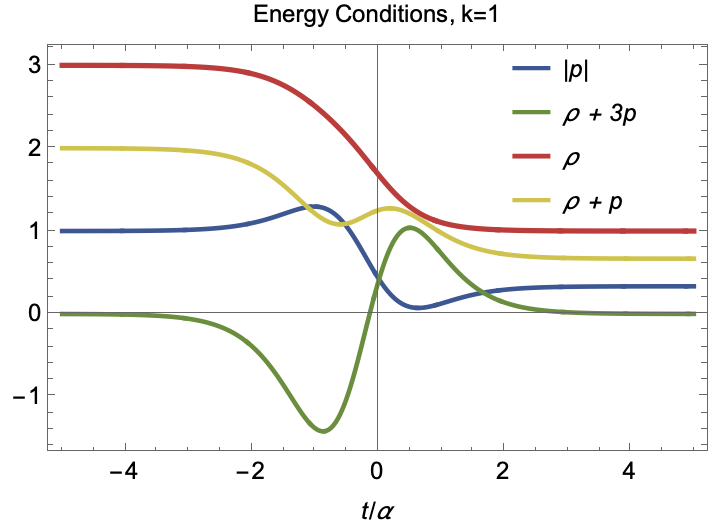}
  \caption{Energy conditions from \eq{atanh} with $k=+1$. Plot of energy density $\rho$ (red), $\rho + p$  (yellow), $|p|$ (blue) and $\rho + 3p$ (green).}
  \label{ECtanhk1}
\end{figure}

Curvature effectively ‘stabilizes’ the energy conditions by contributing a positive term to the ANEC integrand, reinforcing the general result that only closed universes admit nonsingular, geodesically complete FRW cosmologies consistent with the ANEC and supported by physically reasonable matter.

\section{Observational Implications}
Recent large scale structure analyses, including DESI BAO data combined with CMB and supernova constraints, exhibit a mild tendency toward $w(z)< -1$ at redshifts $z\sim0.5-1$ \cite{DESI:2025zgx, Planck:2018vyg}, suggestive of an apparent phantom bias. We emphasize at the outset that we do not interpret this effect as an explanation of
current observational preferences for $w(z)<-1$; rather, it illustrates how even very
small curvature can bias dark-energy reconstructions when flatness is imposed.

Spatial curvature introduces a mild but measurable bias in cosmological parameter inference when data are analyzed under the assumption of exact flatness. In a slightly closed universe ($\Omega_{k,0} < 0$), the curvature contribution to $H(z)$ is absorbed into the dark energy sector when flatness is imposed, producing an effective equation-of-state parameter $w(z) < -1$, i.e., phantom-like behavior \cite{Caldwell:1999ew, Caldwell:2003vq, Carroll:2003st, Vikman:2004dc}. We provide the first explicit analytic derivation of the
curvature-induced phantom bias, consistent with semi-analytic findings~\cite{Clarkson:2007bc}, and embed it within the broader
ANEC-preserving framework of closed cosmologies.

To quantify the effect, consider a true curved $\Lambda$CDM universe,
\begin{equation}
\frac{H^2(z)}{H_0^2} = \Omega_{m,0}(1+z)^3 + \Omega_{\Lambda,0} + \Omega_{k,0}(1+z)^2,
\end{equation}
analyzed using a flat model with the same $\Omega_{m,0}$. The inferred dark energy density in the flat fit is then
\begin{equation}
\rho_{\mathrm{DE}}^{\mathrm{(flat)}}(z) \propto \Omega_{\Lambda,0} + \Omega_{k,0}(1+z)^2,
\end{equation}
normalized to $\rho_{\mathrm{DE}}^{\mathrm{(flat)}}(0) = \Omega_{\Lambda,0} + \Omega_{k,0}$.

The redshift-averaged equation of state over the interval $[0, z]$ is defined by $\rho_{\mathrm{DE}}(z)/\rho_{\mathrm{DE}}(0) = (1+z)^{3(1+w_{\mathrm{eff}})}$, yielding
\begin{equation}
w_{\mathrm{eff}}(z) = -1 + \frac{1}{3} \frac{\ln\left[\Omega_{\Lambda,0} + \Omega_{k,0}(1+z)^2\right] - \ln\left[\Omega_{\Lambda,0} + \Omega_{k,0}\right]}{\ln(1+z)}.
\end{equation}

This form reduces to $w= -1$ in the flat limit and shows that curvature biases $w(z)$ toward more negative values at low redshift.

The local (instantaneous) EOS derived from $d\ln\rho_{\mathrm{DE}}/d\ln(1+z)$ gives
\begin{equation}
w_{\mathrm{inst}}(z) = -1 + \frac{2}{3} \frac{\Omega_{k,0}(1+z)^2}{\Omega_{\Lambda,0} + \Omega_{k,0}(1+z)^2}.
\end{equation}

Both forms predict percent-level shifts in $w(z)$ for $\Omega_{k,0} \sim -0.004$. For instance, at $z = 1$, one finds $w_{\mathrm{eff}} \simeq -1.008$ and $w_{\mathrm{inst}} \simeq -1.016$. Larger phantom deviations (e.g., $w \simeq -1.26$) would require $\Omega_{k,0} \lesssim -0.05$, which is strongly disfavored by current data. Nevertheless, even small curvature-induced biases must be accounted for in precision dark energy analyses.

This small but systematic effect illustrates how even minute departures from exact flatness can manifest observationally as apparent phantom behavior, providing an empirical window on curvature in ANEC-consistent cosmologies.

\section{Conclusions}
We have established a new no-go theorem: spatially flat ($k=0$) and open ($k=-1$) 
Friedmann--Robertson--Walker (FRW) spacetimes cannot be simultaneously nonsingular, 
geodesically complete, and consistent with the averaged null energy condition (ANEC);
any non-static flat or open universe that is complete must therefore violate the ANEC. 
Only ($k=+1$) universes admit such models, with global de~Sitter space providing 
the canonical realization that saturates the bound. Positive spatial curvature thus 
emerges as a fundamental geometric ingredient of nonsingular cosmology.

Our theorem hones the classical singularity results: it is not generic energy 
conditions, but specifically the ANEC, that enforces incompleteness in flat or open 
geometries. Attempts to construct fully regular cosmologies in those spacetimes 
inevitably require ANEC violation---an outcome that is theoretically hazardous, as 
ANEC-violating matter can destabilize the vacuum and permit pathological phenomena 
such as warp drives, traversable wormholes or closed timelike curves. While local NEC violations may arise in effective field theory, consistent ANEC violations along complete, 
achronal null geodesics remain exceedingly difficult to realize.

Closed universes, by contrast, provide a unique geometric loophole. Positive curvature 
contributes to the Raychaudhuri equation like an effective fluid with $w=-1/3$, 
allowing a bounce even in the presence of ordinary matter. Geodesic completeness then 
becomes compatible with both NEC and ANEC, at the mild cost of violating only the 
strong energy condition---a familiar feature of accelerated expansion and inflation.

Taking the ANEC as a fundamental consistency requirement, our results imply that only
closed ($k=+1$) universes can be simultaneously nonsingular and geodesically complete.
This provides a definitive theoretical argument for positive spatial curvature in
cosmology. The universe appears nearly flat ($\Omega_{k,0}\!\simeq\!0$) simply because
it is large—an expected outcome of an inflationary phase. Observationally, even slight
curvature biases flat-model dark-energy reconstructions at the percent level,
underscoring the need to test spatial curvature directly in precision surveys. A
closed primordial universe is consistent with the ANEC and provides the necessary
geometric condition for a fully nonsingular cosmology. Positive curvature is therefore
not merely allowed but preferred on theoretical grounds.

\section*{Acknowledgments}
We thank T. Manton and J.\ Lesnefsky for valuable discussions. DAE is supported in part by the U.S.\ Department of Energy, Office of High Energy Physics, under Award Number DE-SC0019470.

\bibliography{physics}

\end{document}